# Massive scalar counterpart of gravitational waves in scalarized neutron star binaries

Jing Wang[a]

School of Physics and Astronomy, Sun Yat-sen University, Guangzhou 510275, People's Republic of China

**Abstract** In analogy with spontaneous magnetization of ferromagnets below the Curie temperature, a neutron star (NS), with a compactness above a certain critical value, may undergo spontaneous scalarization and exhibit an interior nontrivial scalar configuration. Consequently, the exterior spacetime is changed, and an external scalar field appears, which subsequently triggers a scalarization of its companion. The dynamical interplay produces a gravitational scalar counterpart of tensor gravitational waves. In this paper, we resort to scalar–tensor theory and demonstrate that the gravitational scalar counterpart from a double neutron star (DNS) and a neutron star–white dwarf (NS-WD) system become massive. We report that (1) a gravitational scalar background field, arising from convergence of external scalar fields, plays the role of gravitational scalar counterpart in scalarized DNS binary, and the appearance of a mass-dimensional constant in a Higgs-like gravitational scalar potential is responsible for a massive gravitational scalar counterpart with a mass of the order of the Planck scale; (2) a dipolar gravitational scalar radiated field, resulting from differing binding energies of NS and WD, plays the role of a gravitational scalar counterpart in scalarized orbital shrinking NS-WDs, which oscillates around a local and scalar-energy-density-dependent minimum of the gravitational scalar potential and obtains a mass of the order of about $10^{-21}$ eV/c$^2$.

## 1 Introduction

Although it is clear that Einstein's general relativity has been so far a sound theory in describing the dynamics of neutron star (NS) binary systems, several observations indicated that the orbital decay in the Hulse–Taylor system, PSR 1913+16, is mildly more rapid than that predicted by the general relativistic quadrupole formula [1,2]. The long baseline of precise timing observations for PSR J1738+0333 [3] have also indicated an excess orbital decay, which directly translates to a dipole radiation constraint on deviations from the quadruple formula, according to the Lunar Laser Ranging experiments. It was proposed that this can be relieved by considering that a nontrivial scalar configuration comes about in the strong-field regime [4]. By making an analogy with the spontaneous magnetization of ferromagnets below the Curie temperature, a NS, with a compactness of $\frac{Gm}{Rc^2}$[1] above a certain critical value, will exhibit a nontrivial configuration, and a scalar field settles in the interior [5], i.e. spontaneous scalarization occurs for the NS. The neutron star–white dwarf (NS-WD) binaries usually contain a massive recycled NS [6–8], owing to the recycling process [9], which thus in greater measure tends to undergoing a spontaneous scalarization. It was indicated that NS in a binary pulsar, with a mass of 1.4 $M_\odot$, would develop strong scalar charges even in the absence of external scalar solicitation for strong couplings and with vanishing asymptotic value [5]. The spontaneously scalarized component modifies the exterior spacetime and contributes to an external scalar field $\varphi_{ss}$ around it, which produces a scalar asymptotic solution. In the meantime, a scalarization of NS suffers from a change of compactness [10], which enhances the gravitational interaction with its companion. As a result, the companion star is also scalarized, which is assigned to an induced scalarization [11], and the other external scalar field $\varphi_{is}$ subsequently appears around the secondly scalarized component.

The dynamical interplay between $\varphi_{ss}$ and $\varphi_{is}$ is governed by the following relation [11]:

$$^{(n+1)}\varphi_{ss} = {}^{(0)}\varphi_{ss} + \frac{{}^{(n)}\varphi_{is}}{r},$$
$$^{(n+1)}\varphi_{is} = {}^{(0)}\varphi_{is} + \frac{{}^{(n)}\varphi_{ss}}{r}, \qquad (1)$$

---

[a] e-mail: wangjing6@mail.sysu.edu.cn

[1] $m$ and $R$ are the mass and radius of NS, respectively. $G$ is the Newtonian gravitational constant.

where $^{(0)}\varphi_{ss}$ and $^{(0)}\varphi_{is}$ are the external scalar fields initially produced by the spontaneously scalarized NS "ss" and induced scalarized companion star "is", respectively, $^{(n)}\varphi_{ss}$ and $^{(n)}\varphi_{is}$ represent the $n$th induced external scalar fields around the scalarized components "ss" and "is", respectively, and $r$ is the distance from the center of the binary. The feedback mechanism described by Eq. (1) results in an iteratively induced scalarization of two components, which enhances the strength of external scalar fields, as well as the gravitational interaction between two scalarized stars. Accordingly, the Newtonian gravitational interaction of the binary is modified according to [5]

$$V_{\text{int}} = -G\frac{m_{ss}m_{is}}{R_{ss-is}} - G\frac{\omega_{ss}\omega_{is}}{R_{ss-is}}, \quad (2)$$

where $m_{ss}$ and $m_{is}$ represent the masses of the spontaneously scalarized NS and the induced scalarized companion, $\omega_{ss}$ and $\omega_{is}$ denote the scalar charges of corresponding components with the definition of $\omega_{ss,is} = -\frac{\partial \ln m_{ss,is}(\varphi_{ss,is})}{\partial \varphi_{ss,is}}$ [12], and $R_{ss-is}$ is the orbital separation of the binary. The local Newtonian gravitational constant is accordingly modified as

$$G_{\text{eff}} = G(1 + \omega_{ss}\omega_{is} + \cdots), \quad (3)$$

which is assigned to the effective gravitational constant of the scalarized NS binary system. The second term in the brackets, $\omega_{ss}\omega_{is}$, is for the first- and second-order post-Newtonian corrections to the dynamical system, and the $\cdots$ denotes the terms of dissipative corrections to the Newtonian dynamics that accounts of the backreaction of gravitational-wave emission. Either the continual enhancement of external scalar fields or the different scalar charges carried by two components that sources an emission of dipolar gravitational scalar radiation will contribute to a gravitational scalar counterpart in an in-spiraling NS binary. We assign the in-spiraling NS binaries with both tensor gravitational-wave radiation and dipolar gravitational scalar counterparts to be the scalarized systems. As a consequently, the dynamics of scalarized in-spiraling NS binary is encoded not only by the gravitational tensor metric $g_{\mu\nu}$, but also by a gravitational scalar field $\phi$.

It was shown that "spontaneous scalarization" leads to very significant deviations from Einstein's general relativity in conditions involving binary-pulsar systems [10], which do not necessarily vanish when the weak-field scalar coupling tends to zero. The non-perturbative strong-field deviations away from general relativity due to the appearance of scalar fields, measured by a dimensional scalar coupling factor [10], could have a significant impact on the emission of gravitational waves in NS systems [4]. The equations of motion for scalarized NSs binary systems have been modified, which produce dipolar gravitational scalar counterparts of gravitational tensor waves, depending on the coupling strength between scalar fields and the star matter [11]. In this paper, we resort to the scalar–tensor theory [12] of gravity to describe the dynamics of scalarized in-spiraling NS binary systems and investigate the gravitational scalar counterpart of gravitational waves in scalarized double neutron star (DNS) binaries and NS-WD systems, respectively. We demonstrate that the gravitational scalar counterpart becomes massive in these systems, resulting from different mechanisms. It is pointed out that the appearance of a mass-dimensional constant in the Higgs-like gravitational scalar potential of scalarized DNS systems, arising from the dynamical couplings between the gravitational scalar field $\phi$ and the external scalar fields $\varphi_{ss}$ and $\varphi_{is}$, as well as the self-coupling of $\phi$, contributes to spontaneous symmetry breaking and thus the mass of the gravitational scalar counterpart. During this process, the gravitational scalar fluctuations, because of the iterative interplay between $\varphi_{ss}$ and $\varphi_{is}$, play a role in the Higgs-like field. In NS-WD systems, a monotonically gravitational scalar potential, resulting from the self-couplings of gravitational scalar counterparts, combined with its dynamically couplings to the scalarized stars, makes the gravitational scalar counterparts become massive. We also estimate the masses of the gravitational scalar counterparts in scalarized DNS binaries and scalarized NS-WD systems, respectively. In Sect. 2, we investigate the role of the gravitational scalar counterpart in scalarized DNS binaries and the mechanism that makes it become massive. A different scenario in post-Newtonian corrected in-spiraling scalarized NS-WD binaries is discussed in Sect. 3. Finally, we give a summary in Sect. 4.

## 2 Massive gravitational scalar counterpart of GWs in DNS binary

The feedback effects between $\varphi_{ss}$ and $\varphi_{is}$, described by Eq. (1), contribute to a continuous enhancement of scalar configurations inside two components, as well as the external scalar fields. As a consequence, convergence of $^{(n)}\varphi_{ss}$ and $^{(n)}\varphi_{is}$ occurs, which produces a gravitational scalar background field $\phi_B$. Therefore, the system is immersed in the gravitational scalar background field $\phi_B$, and the dynamics of the DNS binary deviates from Einstein's general relativity, which has influence on its orbital evolution [12–14].

The dynamics of a scalarized in-spiraling DNS binary is then encoded not only by the gravitational tensor metric $g_{\mu\nu}$, but also by a gravitational scalar background field $\phi_B$, which naturally renders the scalar–tensor theory [12] of gravity an alternative to Einstein's general relativity describing the scalarized binary system. Because of a very approximate compactness of the two components in DNS binary, we can neglect the effects of the differences in the couplings between scalar field and the NS matter. Therefore, the scalar–tensor action that describes the scalarized DNS binary can be writ-

ten as

$$S = \int d^4x \sqrt{-g} \left( \frac{M_{\text{pl}}^2}{2} \mathcal{R} - \frac{1}{2} g^{\mu\nu} \partial_\mu \phi_B \partial_\nu \phi_B - V_{\text{DNS}}(\phi_B) \right). \tag{4}$$

Here, $M_{\text{pl}} = \sqrt{1/8\pi G}$ is the reduced Planck constant. $\mathcal{R}$ and $g$ are the Ricci scalar and the determinant of the gravitational tensor metric $g_{\mu\nu}$, respectively. $V_{\text{DNS}}(\phi_B)$ is the gravitational scalar potential of DNS, which consists of the dynamical coupling of $\phi_B$ to $\varphi_{\text{ss,is}}$ and a self-coupling term of $\phi_B$,

$$V_{\text{DNS}}(\phi_B) = \frac{\alpha}{2} \varphi_{\text{ss}} \varphi_{\text{is}} \phi_B^2 + \frac{\lambda}{4} \phi_B^4. \tag{5}$$

Here, $\alpha \equiv -M_{\text{pl}} \frac{d \log m_{\text{ss,is}}}{d \phi_B}$ is a dimensionless coupling constant and characterizes the coupling strength between $\phi_B$ and the matter in the scalarized stars, whose value depends on the compactness of stars consisting of the binary [4,5,12]. $\lambda$ is the self-coupling constant, which is roughly of the order of unity.

The iterative interplay and convergence of $^{(n)}\varphi_{\text{ss,is}}$ perturb $\phi_B$ and cause small gravitational scalar background fluctuations $\sigma$ ($\sigma \ll \phi_B$). The background fluctuating field also has effects on both the gravitational tensor metric and the gravitational background scalar field, via an exponential transformation $e^{\lambda\sigma}$, which follows the couplings [4,10,11]

$$g^*_{\mu\nu} = e^{-2\lambda\sigma} g_{\mu\nu}, \quad \sqrt{-g^*} = e^{4\lambda\sigma} \sqrt{-g}, \tag{6}$$
$$\phi_B^* = e^{-\lambda\sigma} \phi_B, \tag{7}$$

where $g^*_{\mu\nu}$, $g^*$, and $\phi_B^*$ are the transformed gravitational tensor metric and its determinant, and the transformed gravitational background scalar field. By expanding the transformed metric $g^*_{\mu\nu}$ about a Minkowski background in terms of Eq. (6), we express them as

$$g^*_{\mu\nu} = \eta_{\mu\nu} + h^*_{\mu\nu}, \quad h^*_{\mu\nu} = h_{\mu\nu} + 2\eta_{\mu\nu}\lambda\sigma, \tag{8}$$

where $|h_{\mu\nu}|, |h^*_{\mu\nu}| \ll 1$. Equation (6) remains unchanged. Under the transformation of $\sigma$, we find, using Eqs. (6) and (7), that the kinetic term in the action (4) is transformed into a canonical kinetic term,

$$-\frac{1}{2}\sqrt{-g} g^{\mu\nu} \partial_\mu \phi_B \partial_\nu \phi_B = -\frac{1}{2}\sqrt{-g^*} g^{*\mu\nu} \mathcal{D}_\mu \phi_B^* \mathcal{D}_\nu \phi_B^*, \tag{9}$$
$$\mathcal{D}_\mu \equiv \partial_\mu + \lambda \partial_\mu \sigma, \tag{10}$$

which is scale invariant. The transformed action then reads

$$S^* = \int d^4x \sqrt{-g^*} \left( \frac{M_{\text{pl}}^2}{2} \mathcal{R}^* - \frac{1}{2} g^{*\mu\nu} \mathcal{D}_\mu \phi_B^* \mathcal{D}_\nu \phi_B^* \right.$$
$$\left. - V_{\text{DNS}}(\phi_B^*) \right). \tag{11}$$

Now we consider the solution. In the process of performing a conformal transformation, the solutions of external scalar fields $\varphi_{\text{ss}}$ and $\varphi_{\text{is}}$ with mass dimensions [10] involves a dimensional constant $\mu$ with the Planck mass scale, which appears in the transformed gravitational scalar potential $V_{\text{DNS}}(\phi_B^*)$,

$$V_{\text{DNS}}(\phi_B^*) = \frac{\alpha}{2} \mu^2 \phi_B^{*2} + \frac{\lambda}{4} \phi_B^{*4}. \tag{12}$$

The Planck-scale constant $\mu = \sqrt{1/8\pi G_{\text{eff}}}$ here appears to be related to the scalar charges of the scalarized NSs via the effective gravitational constant $G_{\text{eff}}$ according to Eq. (3) [12]. It is the appearance of the mass-dimensional constant $\mu$, which is responsible for a spontaneous breaking of symmetry, which allows us to apply a similar recipe to the Higgs mechanism in the standard model. Thus the gravitational scalar background field becomes a massive one.

Actual NSs observed in DNS binaries, with important deviations from general relativity in the strong-field regime, would develop strong scalar charges in the absence of an external scalar field for sufficiently negative values of $\alpha$, i.e. $\alpha < 0$ [4,5,11]. The self-coupling constant $\lambda$ is of the order of unity, i.e. $\lambda > 0$. By considering that the interplay between $\varphi_{\text{ss}}$ and $\varphi_{\text{is}}$ is a long-range force, the behavior of a transformed gravitational background scalar field $\phi_B^*$ near spatial infinity endows it with a vacuum expectation value (VEV) $v_{\phi_B^*}$,

$$(v_{\phi_B^*})^2 = -\frac{\alpha \mu^2}{2\lambda}, \tag{13}$$

which is obtained from the condition $\frac{dV(\phi_B^*)}{d(\phi_B^*)}|_{(\phi_B^*)_{\min}} = 0$. Therefore, the gravitational scalar background field $\phi_B^*$ is a combination of its VEV $v_{\phi_B^*}$ and the approximate value of the fluctuating field at spatial infinity. Substituting the VEV (13) into the Lagrangian of $\phi_B^*$ extracted from Eq. (11), we get the mass of $\phi_B^*$,

$$(m_s^{\text{DNS}})^2 = -\alpha \mu^2. \tag{14}$$

It was proven that non-perturbative strong-gravitational-field effects develop in NSs for a dimensionless coupling constant $\alpha \lesssim -4$, which causes order-of-unity deviations from general relativity [4]. The general properties of binary systems consisting of scalarized NSs can be described by $\alpha \gtrsim -4.5$, because of binary-pulsar measurements [3,8,15]. For $\alpha \lesssim -5$, NSs in a binary pulsar, with a mass of $1.4 M_\odot$, would develop strong scalar charges even in the absence of external scalar solicitation, and a more negative value of $\alpha$ corresponds to a less compact NS [5]. Most of the measured more massive NSs in detected DNS systems have masses of $\sim 1.3-1.44 M_\odot$ [6]. Consequently, the coupling constant is in the range of $\alpha = -5$ to $-6$ in a quadratic coupling model described in Eq. (12) [5]. The scalar charges mildly vary with the compactness of NSs [11] and will be $\sim 1$ only in the last stages of the evolution of NS binaries or close transient encounters. For NSs in nine so far detected DNS systems, the scalar charges are around 0.2 within the solar-system bound [16] in the Fierz–Jordan–Brans–Dicke theory, by considering its dependence on the "sensitivities" $s \sim 0.2$ [17,18]. Accordingly, the gravitational scalar counterpart of gravitational waves in scalarized in-spiraling DNS binary is of the order of the Planck-mass scale.

## 3 Massive gravitational scalar counterpart of GWs from NS-WD system

It is well known that NS is a more compact object than WD. Consequently, the strength of the coupling between the scalar configuration inside stars and the NS/WD matter is different. A distinct dependence of masses on the scalar fields for NS and WD actually is the source of an emission of dipolar gravitational scalar radiation in a post-Newtonian in-spiraling scalarized binary [11], in addition to the quadruple tensor gravitational waves. Accordingly, the dynamics of a scalarized in-spiraling NS-WD system is governed by a gravitational scalar radiated field $\phi_r$, together with the gravitational tensor metric $g_{\mu\nu}$. The scalar charge of a scalarized NS-WD binary can be extracted from the behavior of the gravitational scalar radiated field near spatial infinity [12], i.e.

$$\phi_r = \phi_r^0 + \frac{\phi_r^1}{r} + \mathcal{O}\left(\frac{1}{r^2}\right), \tag{15}$$

where the iterative interplay and convergence of the external scalar fields $\varphi_{ss}$ and $\varphi_{is}$ around NS and WD are considered, and $\phi_r^0$ is the asymptotic value of the gravitational scalar radiated field at spatial infinity. Accordingly, the dynamics of an in-spiraling scalarized NS-WD binary system, suffering from the post-Newtonian corrections, is described by the following scalar–tensor action:

$$S = \int d^4x \sqrt{-g} \left( \frac{\mathcal{R}}{16\pi G} - \frac{1}{2} g^{\mu\nu} \partial_\mu \phi_r \partial_\nu \phi_r - V_{\text{NS-WD}}(\phi_r) \right) + \sum_n \int_{\gamma_n} ds \, m_{(ss,is)_n}(\phi_r). \tag{16}$$

The gravitational scalar potential of NS-WD binary $V_{\text{NS-WD}}(\phi_r)$ results from two interactions, i.e. the self-interactions of $\phi_r$ and the interactions between $\phi_r$ and matter fields of NS and WD. The gravitational scalar radiated field is associated with the non-perturbative strong-field effects [4], which contributes to the potential of the runaway form [19] that satisfies $\lim_{\phi_r \to \infty} V_{\text{NS-WD}}(\phi_r) \to$

$0$, $\lim_{\phi_r \to \infty} \frac{V_{\text{NS-WD}}(\phi_r)'}{V_{\text{NS-WD}}(\phi_r)} \to 0$, $\lim_{\phi_r \to \infty} \frac{V_{NS-WD}(\phi_r)''}{V_{\text{NS-WD}}(\phi_r)'} \to 0$, ..., as well as $\lim_{\phi_r \to 0} V_{\text{NS-WD}}(\phi_r) \to \infty$, $\lim_{\phi_r \to 0} \frac{V_{\text{NS-WD}}(\phi_r)'}{V_{\text{NS-WD}}(\phi_r)} \to \infty$, $\lim_{\phi_r \to 0} \frac{V_{\text{NS-WD}}(\phi_r)''}{V_{\text{NS-WD}}(\phi_r)'} \to \infty$, ... ($V_{\text{NS-WD}}(\phi_r)' \equiv \frac{dV}{d\phi_r}$, and $V_{\text{NS-WD}}(\phi_r)'' \equiv \frac{d^2V}{d\phi_r^2}$, etc.). Thus, the self-interactions of gravitational scalar radiated field, whose behavior is described by Eq. (15), lead to a monotonically decreasing potential,

$$V_{\phi_r} = \nu^5/\phi_r, \tag{17}$$

where $\nu$ has the unit of mass. The NS/WD matter interacts directly with the gravitational scalar radiated field $\phi_r$ through a conformal coupling of the form $e^{-\alpha_{ss,is}\phi_r/\mu}$. The values of $\alpha_{ss,is}$ are also usually negative for WDs [20]. So the exponential coupling function is an increasing function of $\phi_r$. The combined effects of self-interactions of $\phi_r$ described by Eq. (17) and the conformal coupling give us the form of the scalar potential $V_{\text{NS-WD}}(\phi_r)$ in Eq. (16),

$$V_{\text{NS-WD}}(\phi_r) = \frac{\nu^5}{\phi_r} + \varepsilon_\varphi e^{-\alpha_{ss,is}\phi_r/\mu}. \tag{18}$$

It can be found that $V_{\text{NS-WD}}(\phi_r)$ is an explicit function of the energy density $\varepsilon_\varphi$ of the external scalar fields $\varphi_{ss,is}$, which depends on the masses of the stars (a function of the density for each star $\rho_{ss,is}$) and the coupling strength between interior scalar configuration and matter components of NS/WD [5].

The summation part of Eq. (16) describes the action of the matter components making up the NS and the WD. In the sum over $n$ we give the world line action for any number of species of matter and particles consisting in the NS and the WD and we use $\gamma_n$ to represent the integral of the matter action along the world line. The couplings of matter components inside the stars to the scalar field arise from the dependence of the masses $m_{ss,is}$ on $\phi_r$. The NS/WD matter couples to the gravitational tensor metric $g_{\mu\nu}$ via the conformal transformation $e^{-\alpha_{ss,is}\phi_r/\mu}$, according to the rescaling relation,

$$g^*_{\mu\nu} = e^{-2\alpha_{ss,is}\phi_r/\mu} g_{\mu\nu}. \tag{19}$$

The combined gravitational scalar potential $V_{\text{NS-WD}}(\phi_r)$, Eq. (18) in the NS-WD system, consisting of a monotonically decreasing potential (17) and a monotonically increasing interaction $e^{-\alpha_{ss,is}\phi_r/\mu}$, actually displays a minimum. By minimizing the differentiation of the gravitational scalar potential with respect to $\phi_r$, i.e.

$$V_{\text{NS-WD}}(\phi_r)' - \sum_{ss,is} \frac{\alpha_{ss,is}}{\mu} \varepsilon_\varphi e^{-\alpha_{ss,is}\phi_r/\mu} = 0, \tag{20}$$

we can get the minimum value of $\phi_r$ at the minimum potential $\phi_r^{\min}$. Around this minimum, the gravitational scalar radiated field acquires an effective mass, which is obtained by evaluating the second derivative of the potential at $\phi_r^{\min}$,

$$(m_s^{\text{NS-WD}})^2 = V_{\text{NS-WD}}(\phi_r)''|_{\phi_r^{\min}} + \sum_{\text{ss,is}} \frac{\alpha_{i,j}^2}{\mu^2} \varepsilon_\varphi e^{-\alpha_{\text{ss,is}}\phi_r/\mu}|_{\phi_r^{\min}}. \quad (21)$$

Equations (20) and (21) imply that both the local value of the gravitational scalar radiated field $\phi_r^{\min}$ and the mass of the scalar counterpart depend on the local energy density of external scalar fields produced by two scalarized components. It can be found, from Eq. (21), that the gravitational scalar radiated field becomes more massive in a higher $\varepsilon_\varphi$ environment.

The gravitational scalar interaction between NS and WD, mediated by a massive gravitational scalar radiated field, typically acquires an exponential Yukawa suppression, which results in a finite range of the Yukawa type of potential energy,

$$U(r) = -2\alpha_{\text{ss}}\alpha_{\text{is}} \frac{G m_{\text{ss}} m_{\text{is}}}{r} e^{-m_s^{\text{NS-WD}} r}. \quad (22)$$

Here the product $2\alpha_{\text{ss}}\alpha_{\text{is}}$ is referred to as the interaction strength. The inverse of the range $\lambda$ of a Yukawa potential $e^{-r/\lambda}/r$ characterizes the mass of the gravitational radiated scalar field $\lambda^{-1} \equiv m_s^{\text{NS-WD}}$. Most of the NS-WD binaries have a very small orbital eccentricity, $\sim 10^{-5} - 10^{-6}$ [6], i.e. approximately circular orbits. Accordingly, the scalarized NS and WD orbit with each other and form a ring-configuration-orbit on the binary plane. The distance $a$ from the center of the binary plane to the outer boundary of the ring configuration corresponds to the semi-separation of an NS-WD binary, which is of the order of $\sim 10^9$ m [6], and the central thickness of the ring approximately equals the diameter of the WD, i.e. $\Delta a \sim 10^6$ m. By comparing the radius of NS and WD with the separation between them, we can get $\frac{\Delta a}{a} \sim 10^{-3} \ll 1$. Accordingly, the orbit of the NS-WD binary can be assigned to a thin-ring orbit. The gravitational scalar interactions between NS and WD are therefore screened in the thin-ring configuration, with an interaction range of the same order as the orbital width $\lambda \sim 10^6$ m. Consequently, the corresponding mass of the gravitational radiated scalar field in NS-WD binary is estimated as $m_s^{\text{NS-WD}} \equiv \lambda^{-1} \sim \Delta a^{-1} \sim 10^{-21} eV/c^2$.

## 4 Summary and discussions

In this work, we resort to the scalar–tensor theory of gravity, describing the dynamics of scalarized orbital shrinking NS binary systems, and we investigate the gravitational scalar counterpart of tensor gravitational waves. It was found that a massive NS will develop a nontrivial scalar configuration in the strong-field regime [4], which is consistent with the observations of a relativistic binary-pulsar system, e.g. the DNS system PSR B1913+16 [1,2] and a NS-WD binary PSR J1738+0333 [3]. The spontaneously scalarized NS, with interior scalar configurations, produces a scalar field in its exterior, on the one hand. On the other hand, the external scalar field will change the interactions between two components and the dynamics of the binary, which induces a scalarization of its companion, as well as a second external scalar field around the companion star. The gravitational interaction between the two components is then enhanced, due to the scalar corrections to the Newtonian one (Eq. (2)), which leads to iteratively induced scalarizations. Therefore, the external scalar fields are strengthened continually. Either the interior mass-dependent scalar configuration or the dynamical interplay of external scalar fields causes a gravitational scalar counterpart of quadruple gravitational radiation in in-spiraling NS binaries. Hence, the NS binaries are encoded not only by the tensor metric, but also by a gravitational scalar field, which modifies the dynamics of the binaries and makes the scalar–tensor theory of gravity a natural alternative theory to Einstein's general relativity, describing the in-spiraling systems.

Note that the spontaneous scalarization takes place in the interior of a NS, located in a binary system, when its compactness $\frac{Gm}{Rc^2}$ is above a certain critical threshold even in the absence of scalar sources. The subsequently induced scalarizations also occur in each companion star of the spontaneously scalarized NSs. That is to say, both spontaneous scalarization and induced scalarization occur in the interior of a single star. So the components in DNS binaries and NS-WD systems undergo scalarizations via the same mechanisms. However, the binding energies of NS and WD are different, which contributes to differences in the dependence of the masses of NS/WD on the scalar configurations (i.e. scalar charges). An obvious difference in the dependence of the masses on the scalar field of the two components in one binary system actually is the source of emission of dipolar gravitational scalar radiation [11]. As a result, the causes of the gravitational scalar field $\phi$ are distinct in DNS and NS-WD systems. In in-spiraling DNS binaries, the two NS components possess very similar binding energies [6], and the scalar charges $\omega_{\text{ss}}$ and $\omega_{\text{is}}$ are very close to each other. Accordingly, the dipolar gravitational radiation is negligible. With the iteratively interplay, the strengths of the external scalar field around each component is enhanced, and convergence finally occurs. As a consequence, a gravitational scalar background field appears, which plays the role of the gravitational scalar counterpart of quadruple gravitational tensor waves. In in-spiraling NS-WD systems, owing to a different binding energy of NS and WD, the dependence of masses of NS/WD on the scalar configurations is different. Therefore, the two components in the NS-WD binary carry different scalar charges, which is responsible for the dipolar gravitational scalar radiation. Therefore, the gravitational scalar radiated field plays the role of the gravitational scalar counterpart for quadruple gravitational tensor waves in a post-Newtonian corrected in-spiraling scalarized NS-WD system.

Consequently, the scalarized in-spiraling DNS and NS-WD systems, suffering from gravitational scalar counterparts, are immersed in gravitational scalar potentials, resulting from different mechanisms, which contribute to distinct physical processes. In the in-spiraling scalarized DNS binaries, because of the iterative interplay of two external scalar fields, the gravitational scalar background field suffers from fluctuations. The scalar fluctuations couple to both tensor metrics and gravitational scalar background field, which transfer the couplings of scalar fields into the Higgs-like gravitational scalar potential Eq. (12), with the appearance of a mass-dimensional constant. It is the appearance of the Planck-scale mass-dimensional constant that is responsible for spontaneous breaking of symmetry. Thus the gravitational scalar background field becomes a massive one, in which the gravitational scalar fluctuation field is the massless field and plays the role of a Higgs-like field. Therefore, the mass of the gravitational scalar counterpart in in-spiraling scalarized DNS, expressed as Eq. (14), is of the order of the Planck mass scale, which depends on the coupling strength between the gravitational background scalar field and NS matter. In in-spiraling scalarized NS-WD binaries, the gravitational scalar potential is then consisting of a monotonically decreasing self-interaction of the gravitational scalar radiated field and an scalar-energy-density-dependent exponential increasing the coupling to the NS/WD matter. The non-monotonical potential displays a minimum, which contributes to a massive gravitational scalar counterpart. The reason why the gravitational scalar counterpart in the NS-WD system becomes massive is that the gravitational scalar radiated field oscillates around a local minimum of the gravitational scalar potential, with high scalar-energy density. By considering the Yukawa-suppression effects on an environment of high scalar-energy density, we estimate the mass of the dipolar gravitational scalar counterpart of quadruple tensor gravitational waves in NS-WD binaries, expressed by Eq. (21), to be of the order of $\sim 10^{-21}$ eV/c$^2$, which depends on the orbital scale of the binary.

The gravitational waves radiated from in-spiraling DNS and NS-WD binaries, which possess a wide separation with orbital periods in units of days, are located in the typical low-frequency band of around $10^{-4}$ Hz. The amplitudes are of the order of $10^{-24}$. So it is very unlikely that one will be able to detect this currently by LIGO. However, the first space-based gravitational-wave observatory, LISA, is expect to detect space-born low-frequency gravitational waves, whose sensitivity can be reduced to $10^{-24}$. Therefore, we would expect the gravitational waves from in-spiraling scalarized DNS and NS-WD binaries and the scalar counterparts to be detected and constrained potentially by LISA/eLISA in the near future. Currently, we just can expect and try to constrain our results from binary-pulsar observations, which is work under way.

**Acknowledgements** This work is supported by the Fundamental Research Funds for the Central Universities (Grant no. 161gpy49) at Sun Yat-sen University and the Science and Technology Program of Guangzhou (71000-42050001).

### References

1. J.H. Taylor, J.M. Weisberg, A new test of general relativity—gravitational radiation and the binary pulsar PSR 1913+16. Astrophys. J. **253**, 908 (1982)
2. J.M. Weisberg, D.J. Nice, J.H. Taylor, Timing measurements of the relativistic binary pulsar PSR B1913+16. Astrophys. J. **722**, 1030 (2010)
3. P.C.C. Freire, N. Wex, G. Esposito-Farèse, J.P.W. Verbiest, M. Bailes, B.A. Jacoby, M. Kramer, I.H. Stairs, J. Antoniadis, G.H. Janssen, The relativistic pulsar-white dwarf binary PSR J1738+0333—II. The most stringent test of scalar–tensor gravity. Mon. Not. R. Astron. Soc. **423**, 3328 (2012)
4. T. Damour, G. Esposito-Farese, Nonperturbative strong field effects in tensor–scalar theories of gravitation. Phys. Rev. Lett. **70**, 2220 (1993)
5. T. Damour, G. Esposito-Farese, Tensor–scalar gravity and binary pulsar experiments. Phys. Rev. D **54**, 1474 (1996)
6. C.M. Zhang, J. Wang, Y.H. Zhao, H.X. Yin, L.M. Song, D.P. Menezes, D.T. Wickramasinghe, L. Ferrario, P. Chardonnet, Study of measured pulsar masses and their possible conclusions. Astron. Astrophys. **527**, A83 (2011)
7. P.B. Demorest, T. Pennucci, S.M. Ransom, M.S.E. Roberts, J.W.T. Hessels, A two-solar-mass neutron star measured using Shapiro delay. Nature **467**, 1081 (2010)
8. J. Antoniadis, P.C.C. Freire, N. Wex, T.M. Tauris, R.S. Lynch, M.H. van Kerkwijk, M. Kramer, C. Bassa, V.S. Dhillon, T. Driebe, J.W.T. Hessels, V.M. Kaspi, V.I. Kondratiev, N. Langer, T.R. Marsh, M.A. McLaughlin, T.T. Pennucci, S.M. Ransom, I.H. Stairs, J. van Leeuwen, J.P.W. Verbiest, D.G. Whelan, A massive pulsar in a compact relativistic binary. Science **340**, 448 (2013)
9. J. Wang, C.M. Zhang, Y.H. Zhao, Y. Kojima, H.X. Yin, L.M. Song, Spin period evolution of a recycled pulsar in an accreting binary. Astron. Astrophys. **526**, A88 (2011)
10. M. Salgado, D. Sudarsky, U. Nucamendi, On spontaneous scalarization. Phys. Rev. D **58**, 124003 (1998)
11. C. Palenzuela, E. Barausse, M. Ponce, L. Lehner, Dynamical scalarization of neutron stars in scalar–tensor gravity theories. Phys. Rev. D **89**(4), 044024 (2014)
12. T. Damour, G. Esposito-Farese, Tensor multiscalar theories of gravitation. Class. Quant. Grav. **9**, 2093 (1992)

Eur. Phys. J. C (2017) 77:641	Page 7 of 7	64113. R.A. Hulse, J.H. Taylor, Discovery of a pulsar in a binary system. Astrophys. J. Lett. **195**, L51 (1975)
14. C.M. Will, H.W. Zaglauer, Gravitational radiation, close binary systems, and the Brans–Dicke theory of gravity. Astrophys. J **346**, 366 (1989)
15. T. Damour, G. Esposito-Farese, Gravitational wave versus binary-pulsar tests of strong field gravity. Phys. Rev. D **58**, 042001 (1998)
16. D.M. Eardley, Observable effects of a scalar gravitational field in a binary pulsar. Astrophys. J. Lett. **196**, L59 (1975)
17. T. Damour, G. Esposito-Farèse, Testing gravity to second post-Newtonian order: a field theory approach. Phys. Rev. D **53**, 5541 (1996)
18. S. Mirshekari, C.M. Will, Compact binary systems in scalar–tensor gravity: equations of motion to 2.5 post-Newtonian order. Phys. Rev. D **87**, 084070 (2013)
19. J. Khoury, A. Weltman, Chameleon cosmology. Phys. Rev. D **69**, 044026 (2004)
20. T. Damour, G.W. Gibbons, C. Gundlach, Dark matter, time varying $G$, and a dilaton field. Phys. Rev. Lett. **64**, 123 (1990)Springer